# The Planck distribution of phonons in a Bose-Einstein condensate

R. Schley,[1] A. Berkovitz,[1] S. Rinott,[1] I. Shammass,[2] A. Blumkin,[1] and J. Steinhauer[1]

[1]Department of Physics, Technion—Israel Institute of Technology, Technion City, Haifa 32000, Israel

[2]Department of Condensed Matter Physics, Weizmann Institute of Science, Rehovot 76100, Israel

The Planck distribution of photons emitted by a black body led to the development of quantum theory. An analogous distribution of phonons should exist in a Bose-Einstein condensate. We observe this Planck distribution of thermal phonons in a 3D condensate. This observation provides an important confirmation of the basic nature of the condensate's quantized excitations. In contrast to the bunching effect, the density fluctuations are seen to increase with increasing temperature. This is due to the non-conservation of the number of phonons. In the case of rapid cooling, the phonon temperature is out of equilibrium with the surrounding thermal cloud. In this case, a Bose-Einstein condensate is not as cold as previously thought. These measurements are enabled by our *in situ k*-space technique.

Quantum theory was first discovered by considering the spectrum of light emitted by a black body [1]. The correct spectrum was only obtained when it was realized that light is quantized into photons. The population of these photons is given by the Planck distribution, which increases with the temperature of the black body. A Planck distribution of phonons was then used to explain the specific heat of crystals. This was an important early success of quantum theory. It is also believed that a Planck distribution of phonons should occur in a 3D Bose-Einstein condensate [2,3]. By employing our new *in situ* $k$-space technique, we make the first observation of the Planck distribution of these naturally-occurring phonons.

Large quantities of artificially created phonons have been studied previously by Bragg scattering [4-8]. However, this technique does not measure the population of phonons in the condensate. Indeed, Ref. 5 presented the future goal of observing the thermal phonons. We have now reached this goal. The Bragg scattering measurements have included the phonon energy [5], the zero-temperature static structure factor [5] and the Bogoliubov amplitudes [7]. The full dispersion relation and $k$-dependence of the zero-temperature static structure factor were also measured [6,9], and the current work will rely on these relations. All of these measurements involved creating and observing large numbers of phonons. In contrast, we observe small populations of spontaneously occurring phonons, which arise due to the finite temperature of the condensate $T_{\text{cond}}$. Due to this temperature, the population of the phonons should have the Planck distribution $N_{\mathbf{k}} = \left(e^{\hbar\omega_k/k_B T_{\text{cond}}} - 1\right)^{-1}$ [2,3], where $\hbar\omega_k$ is the excitation energy at finite temperature [10].

The thermal phonons result in density fluctuations, as quantified by the static structure factor, which can be written as [11]

$$S(k) = \frac{1}{N}[\langle|\rho_{\mathbf{k}}|^2\rangle - |\langle\rho_{\mathbf{k}}\rangle|^2] \tag{1}$$

where $\rho_{\mathbf{k}}$ is the Fourier transform of the density $\rho(\mathbf{r})$, and $N$ is the total number of atoms. $S(k)$ gives the ensemble average of the density fluctuations with wave number $k$ [12]. For a homogeneous condensate at finite temperature, $S(k)$ is given by [5,11]

$$S(k)_{\text{cond}} = \left(N_{\mathbf{k}} + N_{-\mathbf{k}} + 1\right)S_o(k) \tag{2}$$

where $S_o(k) = (u_k + v_k)^2$ is the zero temperature static structure factor, $u_k$ and $v_k$ are the Bogoliubov amplitudes, and $N_{-\mathbf{k}} = N_\mathbf{k}$. Even at zero temperature where $N_{\pm\mathbf{k}} = 0$, the interacting condensate contains atoms with non-zero momenta. These cause the unity term in parentheses, the quantum fluctuations. For the rest of this work it is understood that (2) is averaged in the local density approximation, to account for the inhomogeneous density in the harmonic trap. Expression (2) also applies to 1D and 2D Bose gases, and it has been seen to result in number fluctuations in small subvolumes of these gases [13,14].

Our observation of the Planck distribution of phonons is not related to the bunching phenomenon, since a 3D condensate has no bunching [15,16]. However, we must also account for the bunching fluctuations $S(k)_{\text{therm}}$ resulting from the non-condensed atoms present in the atomic sample, particularly at the higher temperatures studied. These additional fluctuations are analogous to the bunching and anti-bunching observed for a 3D Bose gas [15-17], a 1D Bose gas [18,19], a Fermi gas [17,20-22], and a Mott insulator [23,24]. The uncondensed gas is similar to an ideal bose gas, which consists of atomic populations with the Bose distribution $n_\mathbf{k} = \left[e^{(\hbar^2 k^2/2m - \mu)/k_B T_{\text{therm}}} - 1\right]^{-1}$, where $T_{\text{therm}}$ is the temperature of the thermal gas, $\mu$ is the chemical potential, and $m$ is the atomic mass. This results in density fluctuations given by [1,25]

$$S(k)_{\text{therm}} = 1 + \frac{1}{(2\pi)^3 n} \int n_{\mathbf{k}'} n_{\mathbf{k}'+\mathbf{k}} d^3 k' \qquad (3)$$

where $n$ is the density. Due to the conserved number of atoms, the long wavelength $n_\mathbf{k}$ decrease for increasing temperature. Thus, the bunching fluctuations of (3) *decrease* for increasing temperature. It is interesting to contrast these bunching fluctuations with those due to phonons. The phonons are Bogoliubov excitations on a background condensate, and their number is not conserved. Indeed, the Planck distribution is an increasing function of temperature for every wavenumber $k$. Thus, the density fluctuations for the phonons *increase* with increasing temperature.

In order to account for the bunching fluctuations in addition to the phonons, we make the standard two-fluid [3] or bimodal [26,27] approximation, in which the system is approximated as two separate entities: A condensate containing $N_c$ atoms with phonon temperature $T_{\text{cond}}$, and a non-condensed thermal cloud containing $N_T$ atoms at temperature $T_{\text{therm}}$. This approximation allows us to avoid calculating the full spectrum of excitations, and their associated density fluctuations, for the inhomogeneous gas. The condensate and thermal clouds are spatially separated to a large degree, firstly because the thermal cloud primarily consists of high-energy excitations, which are localized at larger radii in the harmonic trap than is the condensate. These excitations have energies greater than $\mu$, and $k$ greater than $\xi^{-1}$, where $\xi$ is the healing length. Secondly, the condensate repels the thermal atoms due to the exchange symmetry [11]. Since the fluctuations of the condensate and thermal cloud are independent, the combined static structure factor is

$$S(k) = \frac{N_c}{N} S(k)_{\text{cond}} + \frac{N_T}{N} S(k)_{\text{therm}}. \qquad (4)$$

We measure $S(k)$ via (1), in order to study the phonon population. We obtain an ensemble of *in situ* images, and average the square of the magnitude of their Fourier transforms, in order to compute the first term in brackets. We also compute the second term, which removes the unwanted contribution from the overall density profile of the inhomogeneous cloud. This is in contrast to previous experiments which measured the zero-temperature static structure factor by observing the response of the gas to a Bragg pulse [5,6], a technique which is insensitive to the phonon population [5].

There are two advantages to detecting phonons *in situ* rather than outside of the condensate in time-of-flight. Firstly, the excited component of the wavefunction interferes with the large condensate, which enhances the measurement by a factor of $\sqrt{N_c}$. For our condensate of $1.3 \times 10^5$ atoms, this gives two orders of magnitude more sensitivity than if we measured the phonons outside of the condensate. Secondly, the Fourier-transform measurement is made at a high spatial frequency $k$, which reduces the noise. However, the Fourier transform technique is only applicable for $k$-values which are resolved by the imaging system. Thus, we rely on the very high spatial resolution of our apparatus [28]. The measured response of the imaging system is shown in Fig. 1(g) [9,25]. This response was obtained by creating and imaging phonons for each value of $k$. It is seen in the figure that much of the phonon regime ($k \leq \xi^{-1}$) is resolved, in that the response is on the order of unity.

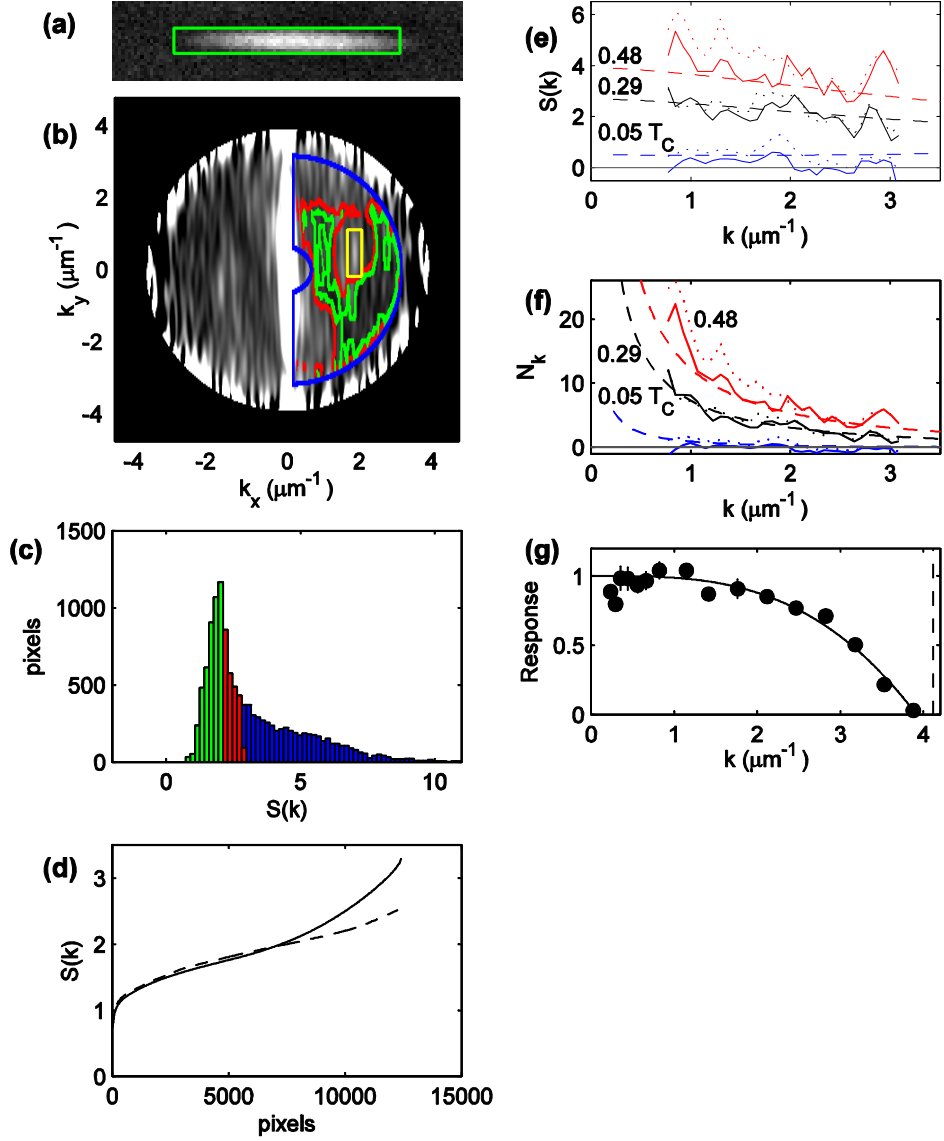

FIG. 1. The measured Planck distribution of phonons. (a) *In situ* image of the atomic cloud at $T_{\text{therm}} = 0.05\, T_c$. The green rectangle indicates the region used for the Fourier transform. The area of the image is 73 μm × 16 μm. (b) The static structure factor in the $k_z = 0$ plane, for $T_{\text{therm}} = 0.05\, T_c$. The area within the green (red) curve is the minimal (maximal) clean window used to compute $S(k)$. The yellow rectangle indicates one of a pair of symmetric peaks, corresponding to an imaging fringe in (a), which is excluded from the clean windows. (c) The histogram of the pixels within the blue curve of (b), computed for the average over all temperatures. The green (red+green) region corresponds to the area within the green (red) curve of (b). (d) The median and mean of the histogram, indicated by dashed and solid curves, respectively. (e) $S(k)$, found by averaging over the polar angle within the clean windows of (b). The blue, black, and red curves correspond to $T_{therm} = 0.05\, T_c$, $0.29\, T_c$, and $0.48\, T_c$, respectively. The solid (dotted) curves correspond to the minimal (maximal) clean windows. The dashed curves show the theoretically expected values, given by (4) with $T_{cond} = T_{therm}$. (f) The phonon population $N_k$. The dashed curves indicate the expected Planck distribution with temperature $T_{cond} = T_{therm}$. In order to extract $N_k$ from $S(k)$, $S(k)_{cond}$ must first be determined. This is achieved by subtracting the contribution due to $S(k)_{therm}$ in (4), by means of the theoretical expression (3). This contribution is significant for the highest temperature only. $N_k$ is then determined by (2). (g) The measured response of the imaging system. The solid curve is a polynomial fit, which is taken as the response function in the analysis of the images. The dashed line indicates $\xi^{-1}$.

The atomic cloud of $^{87}$Rb atoms in the $F = 2$, $m_F = 2$ state is confined in a harmonic magnetic trap with radial and axial frequencies of 224 Hz and 26 Hz, respectively. We image an ensemble of between 20 and 70 clouds *in situ*, as shown in Fig. 1(a). Phase-contrast imaging is employed, with a short 2 µs imaging pulse. The temperature $T_{therm}$ is adjusted by varying the final RF evaporation frequency. For temperatures with a significant condensate fraction, the imaging beam is detuned by 1.3 GHz. This relatively large detuning is used to avoid perturbing the measurement of the dense cloud. For temperatures corresponding to a mostly thermal cloud, a smaller detuning of 210 MHz is used, in order to increase the signal.

The temperature $T_{\text{therm}}$ is determined by fitting a bimodal density profile to the image of Fig. 1(a). The observed critical temperature $T_c$ is 390 nK. In the bimodal fit, the Thomas-Fermi approximation is used for the condensate density, and the thermal density is obtained from a semiclassical approximation to the Hartree-Fock equations [11]. For the lowest temperatures, the noise in the bimodal fit increases. Thus, we use a linear fit to the temperature versus the final evaporation frequency for $T_{\text{therm}} \leq 0.48\, T_c$.

The technique of analyzing the images is essentially as in [25]. We compute the right side of (1) using a 2D Fourier transform within the green rectangle of Fig. 1(a). We then subtract off the measured shot noise, and divide by the square of the response function of the imaging system shown in Fig. 1(g). Fig. 1(b) shows the resulting $S(k_x, k_y, k_z = 0)$. The area outlined in blue avoids the central vertical strip corresponding to the overall profile of the atomic cloud, while remaining within the resolution of the imaging system. This blue region contains $k$-values ranging from 0.7 μm$^{-1}$ to 3.1 μm$^{-1}$. The blue region also contains artifacts: pairs of symmetric peaks in Fig. 1(b) corresponding to sinusoidal fringes in the image. The yellow rectangle indicates one peak of such a pair. While these peaks can be removed by eye, we remove them by averaging Fig. 1(b) over all temperatures, and studying the histogram of the pixels in the blue region, Fig. 1(c). The "tail" in the histogram extending to the right corresponds to the imaging fringes. One way of removing this tail is to consider the left peak of the histogram only, as indicated in green. This gives a minimal estimate for the "clean window", as outlined in green in Fig. 1(b). On the other hand, one can assume that the distribution of random noise in the atomic fluctuations is symmetric, so the mean of the distribution should be equal to the median. This is only the case when the number of pixels is limited to approximately 7000, as shown in Fig. 1(d). This thus gives a maximal estimate of the clean window, as indicated by red+green in Fig. 1(c), and outlined in red in Fig 1(b). The maximal clean window is larger than the minimal clean window by a factor of 1.5.

$S(k)$ is found by an average over the angle in the $k_x$-$k_y$ plane, within the clean window. The solid and dotted curves of Fig. 1(e) show this measured $S(k)$ at various temperatures, for the minimal and maximal clean windows, respectively. Good agreement is seen with the model (4). The corresponding phonon population $N_\mathbf{k}$ is shown in Fig. 1(f), with no free parameters. This is the measured Planck distribution. Good quantitative agreement is seen with the theoretical Planck distribution. The phonon population is seen to increase greatly as the temperature increases, indicating the non-conservation of the phonon number.

By averaging Fig. 1(e) over $k$, $\bar{S}(k)$ is obtained, as shown in Fig. 2(a) as a function of $T_{\text{therm}}/T_c$. It is seen that the results do not depend significantly on which of the clean windows is used. As $T_{\text{therm}}$ decreases toward zero, the fluctuations are also seen to go toward zero, indicating that the phonons "freeze out". This can also be seen in the corresponding $\bar{N}_\mathbf{k}$ (Fig. 3(a), open circles), and in the corresponding phonon temperature $T_{\text{cond}}$ (Fig. 3(b), open circles). Previously, thermometry was achieved in a Fermi gas by studying antibunching [21,22]. For increasing temperatures, $\bar{S}(k)$ reaches a maximum and then decreases as $T_c$ is approached. This decrease is due to the increasing thermal fraction, which has smaller fluctuations than the condensate. Above $T_c$, $\bar{S}(k)$ decreases toward unity [25].

The precision of the measurement is seen to be less than one phonon quantum per mode, for the lower temperatures in Fig. 3(a). For the lowest temperature point, $\bar{S}(k)$ in Fig. 2(a) approaches the quantum fluctuations indicated by the dashed curve. This implies that the unity term in (2) dominates the $N_{\pm \mathbf{k}}$ terms. The measured population even reaches the dotted curve in Fig. 3a, which is the population of phonons expected in analog Hawking radiation [29-32] for a sonic black hole with the maximum possible Hawking temperature $\mu/\pi$ [30].

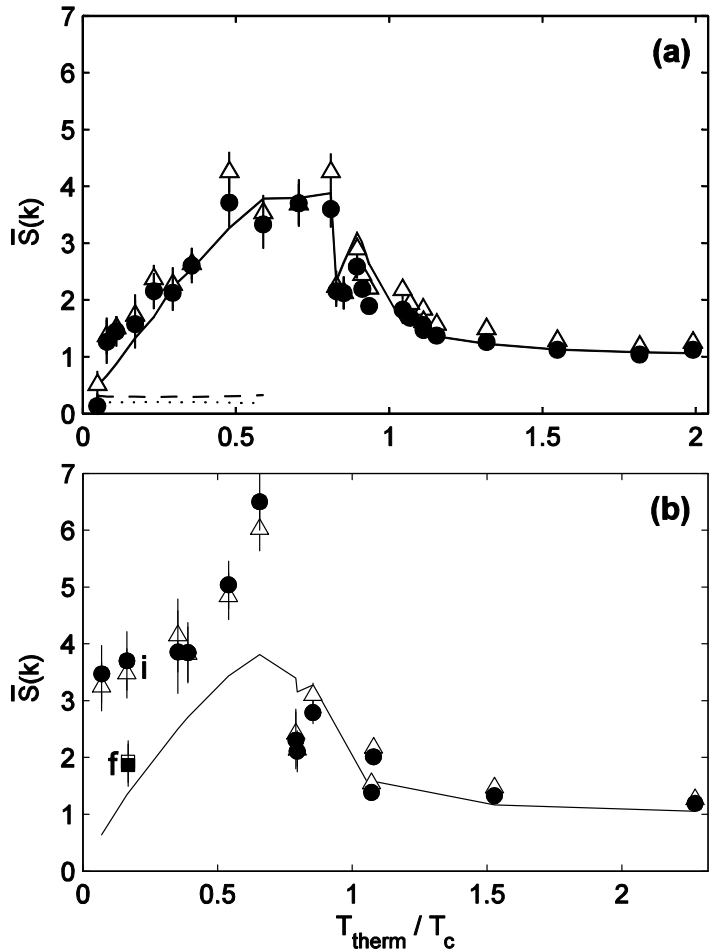

FIG. 2. The static structure factor as a function of temperature. There are no free parameters in the measured points or in the model (solid curve). The circles and triangles are computed with the minimal and maximal clean windows, respectively. The error bars throughout this work indicate the standard error of the mean. (a) Slow cooling. The dashed curve indicates the theoretical quantum fluctuations. The dotted curve indicates the maximum possible Hawking radiation for a sonic black hole. The data above $T_c$ also appeared in [25]. (b) Rapid cooling, resulting in excess fluctuations. The fluctuations at the point labeled "i" decay to the square labeled "f" in 700 ms. It is seen that the system has largely equilibrated during this time, in that the solid curve is almost reached.

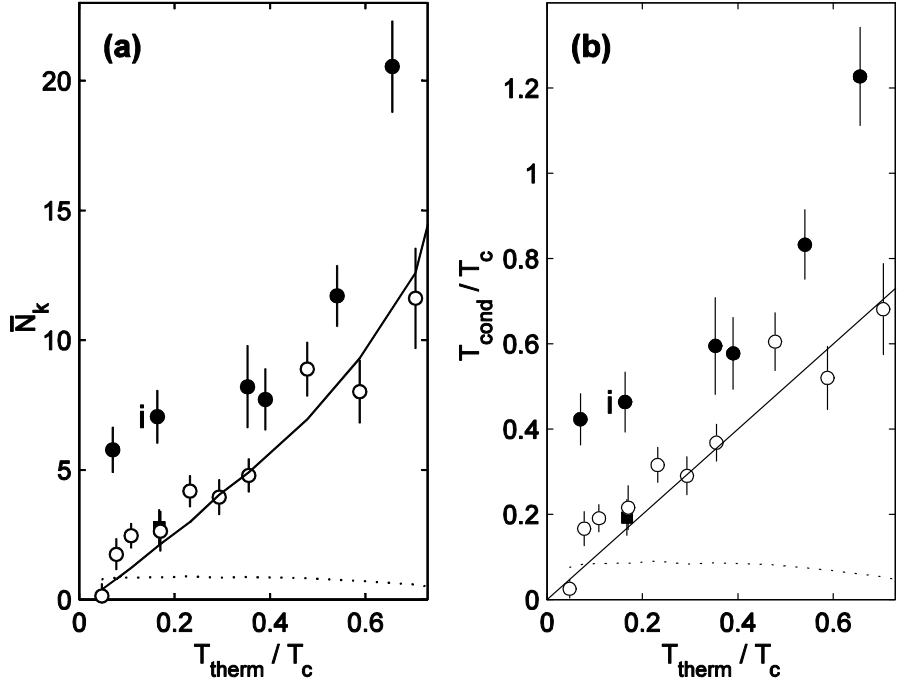

FIG. 3. The average phonon population (a) and the phonon temperature (b). Slow cooling is indicated by open circles, and rapid cooling by filled circles. Each point is an average of the values for the maximal and minimal clean windows. The point "i" for rapid cooling decays to the square in 700 ms. The dotted curve indicates the maximum possible Hawking radiation. The solid line indicates thermal equilibrium ($T_{\mathrm{cond}} = T_{\mathrm{therm}}$). In (a), this curve is found by averaging over the Planck distribution for temperature $T_{\mathrm{therm}}$.

The solid curve in Fig. 2(a) shows the theoretical model (4), with no free parameters. This curve is found by assuming thermal equilibrium ($T_{\mathrm{cond}} = T_{\mathrm{therm}}$). The quantities $N_c/N$ and $N_T/N$ are taken from the bimodal fit. These ratios are evaluated for the volume enclosed within the green rectangle of Fig. 1(a). Good agreement is seen between the model and the measured values at all temperatures, including the regime close to $T_{\text{therm}} = T_c$. Each theoretical point in Fig. 2 is calculated for the measured populations, which results in the jagged appearance of the theoretical solid curves.

The thermal equilibrium seen in Fig. 2(a) relies upon a sufficiently slow cooling rate. We have also increased the sweep rate during the last stage of evaporation (the stage which includes the condensate creation) by a factor of 4 relative to that of Fig. 2(a). Fig. 2(b) shows that this faster cooling indeed results in more phonons, since $\bar{S}(k)$ is well above the theoretical thermal equilibrium curve, for the values of $T_{\text{therm}}$ with a substantial condensate fraction. This can also be seen in the filled circles of Figs. 3(a) and 3(b). We also verify that thermal equilibrium is eventually reached after the rapid evaporation. We start with a condensate with excess phonons immediately after the rapid evaporation sweep, as indicated by the point marked "i" in Figs. 2(b), 3(a), and 3(b). $T_{\text{cond}}$ is significantly greater than $T_{\text{therm}}$ at this time, as seen in Fig. 3(b). We then wait 700 ms between the end of the evaporation sweep, and the imaging of the condensate. An RF shield is present during this wait period. The square in Figs. 2(b), 3(a), and 3(b) indicates the resulting fluctuations. It is seen that the phonons have decayed and the system has largely equilibrated, in that the solid curve is almost reached.

There are two factors which should result in the excess phonons seen for rapid cooling. Firstly, the spatial separation between the phonons and the high-energy excitations prevents equilibration as the cooling continues below $T_c$. Secondly, the Kibble-Zurek mechanism [33-39] should add energy to the condensate as $T_c$ is crossed. This mechanism limits the correlation length $\hat{\xi}$ at the phase transition. The kinetic energy density is proportional to $|\nabla\psi|^2$ [11] and therefore scales as $\hat{\xi}^{-2}$. This scaling includes the vortex case [33,35,37]. We suggest that the Kibble-Zurek mechanism should also create phonons.

In conclusion, we have measured the Planck distribution of phonons in a 3D condensate, as a function of temperature. For slow cooling, the phonon population is close to thermal equilibrium, in analogy with black body radiation. It is seen that the phonon number is not conserved. For sufficiently low temperatures, the phonon population is so small that quantum fluctuations dominate. For rapid cooling however, we see that the phonon population is out of equilibrium with the high-energy excitations, and the condensate contains copious additional phonons. This phenomenon is likely due to the spatial separation between the condensate and thermal clouds, as well as the Kibble-Zurek mechanism. In the case of slow cooling, it is found that a phonon state can be prepared and detected, whose population is sufficiently small for the study of analogue Hawking radiation in a sonic black hole experiment. Our powerful $k$-space technique could be utilized in a variety of additional studies, such as phase transitions in lower dimensions.

We thank N. Pavloff, L. I. Glazman, L. P. Pitaevskii, G. V. Shlyapnikov, R. J. Rivers, I. Zapata, R. Pugatch, D. Podolsky, J. R. Anglin, B. Damski, W. H. Zurek, E. Polturak, Y. Kafri, and B. Shapiro for helpful conversations. This work was supported by the Israel Science Foundation.

[1] L. D. Landau and E. M. Lifshitz, *Course of Theoretical Physics, Volume 5, Statistical Physics, Part 1* (Pergamon Press, Oxford, 1991), Sections 63, 64, 71, and 117.

[2] K. Huang, *Statistical Mechanics* (Wiley, New York, 1987), Chap. 12.

[3] Ph. Nozières, Ph. and D. Pines, *The Theory of Quantum Liquids* (Addison-Wesley, Reading, MA, 1990), Vol. II, Chap. 1 and 2.

[4] R. Ozeri, N. Katz, J. Steinhauer, and N. Davidson, Rev. Mod. Phys. **77**, 187 (2005).

[5] D. M. Stamper-Kurn, A. P. Chikkatur, A. Görlitz, S. Inouye, S. Gupta, D. E. Pritchard, and W. Ketterle, Phys. Rev. Lett. **83**, 2876 (1999).

[6] J. Steinhauer, R. Ozeri, N. Katz, and N. Davidson, Phys. Rev. Lett. **88**, 120407 (2002).

[7]. J. M. Vogels, K. Xu, C. Raman, J. R. Abo-Shaeer, and W. Ketterle, Phys. Rev. Lett. **88**, 060402 (2002).

[8] J. M. Pino, R. J. Wild, P. Makotyn, D. S. Jin, and E. A. Cornell, Phys. Rev. A **83**, 033615 (2011).

[9] I. Shammass, S. Rinott, A. Berkovitz, R. Schley, and J. Steinhauer, Phys. Rev. Lett. **109**, 195301 (2012).

[10] The phonon energy at finite temperature is obtained from the Hartree-Fock-Bogoliubov equations in the Popov approximation [11], yielding the Bogoliubov spectrum with no energy shift due to the density of thermal atoms (see H. Shi, and A. Griffin, Phys. Rep. **304**, 1 (1998)).

[11] L. Pitaevskii and S. Stringari, *Bose-Einstein Condensation* (Oxford University Press, Oxford, 2003), Chap. 7 and 13.

[12] D. Pines and Ph. Nozières, *The Theory of Quantum Liquids* (Addison-Wesley, Reading, MA, 1988), Vol. I, Chap. 2.

[13] C.-L. Hung, X. Zhang, L.-C. Ha, S.-K. Tung, N. Gemelke, and C. Chin, New. J. Phys. **13**, 075019 (2011).

[14] J. Armijo, Phys. Rev. Lett. **108**, 225306 (2012).

[15] E. A. Burt, R. W. Ghrist, C. J. Myatt, M. J. Holland, E. A. Cornell, and C. E. Wieman, Phys. Rev. Lett. **79**, 337 (1997).

[16] M. Schellekens, R. Hoppeler, A. Perrin, J. Viana Gomes, D. Boiron, A. Aspect, and C. I. Westbrook, Science **310**, 648 (2005).

[17] T. Jeltes, J. M. McNamara, W. Hogervorst, W. Vassen, V. Krachmalnicoff, M. Schellekens, A. Perrin, H. Chang, D. Boiron, A. Aspect, and C. I. Westbrook, Nature **445**, 402 (2007).

[18] J. Esteve, J.-B. Trebbia, T. Schumm, A. Aspect, C. I. Westbrook, and I. Bouchoule, Phys. Rev. Lett. **96**, 130403 (2006).

[19] T. Jacqmin, J. Armijo, T. Berrada, K. V. Kheruntsyan, and I. Bouchoule, Phys. Rev. Lett. **106**, 230405 (2011).

[20] T. Rom, Th. Best, D. van Oosten, U. Schneider, S. Fölling, B. Paredes, and I. Bloch, Nature **444**, 733 (2006).

[21] T. Müller, B. Zimmermann, J. Meineke, J.-P. Brantut, T. Esslinger, and H. Moritz, Phys. Rev. Lett. **105**, 040401 (2010).

[22] C. Sanner, E. J. Su, A. Keshet, R. Gommers, Y. Shin, W. Huang, and W. Ketterle, Phys. Rev. Lett. **105**, 040402 (2010).

[23] S. Fölling, F. Gerbier, A. Widera, O. Mandel, T. Gericke, and I. Bloch, Nature **434**, 481 (2005).

[24] N. Gemelke, X. Zhang, C.-L. Hung and C. Chin, Nature **460**, 995 (2009).

[25] A. Blumkin, S. Rinott, R. Schley, A. Berkovitz, I. Shammass, and J. Steinhauer, to be published.

[26] M. H. Anderson, J. R. Ensher, M. R. Matthews, C. E. Wieman, and E. A. Cornell, Science **269**, 198 (1995).

[27] K. B. Davis, M.-O. Mewes, M. R. Andrews, N. J. van Druten, D. S. Durfee, D. Kurn, and W. Ketterle, Phys. Rev. Lett. **75**, 3969 (1995).

[28] S. Levy., E. Lahoud, I. Shomroni, and J. Steinhauer, Nature **449**, 579 (2007).

[29] W. G. Unruh, Phys. Rev. Lett. **46**, 1351 (1981).

[30] O. Lahav, A. Itah, A. Blumkin, C. Gordon, S. Rinott, A. Zayats, and J. Steinhauer, Phys. Rev. Lett. **105**, 240401 (2010).

[31] L. J. Garay, J. R. Anglin, J. I. Cirac, and P. Zoller, Phys. Rev. Lett. **85**, 4643 (2000).

[32] S. W. Hawking, Nature **248**, 30 (1974).

[33] W. H. Zurek, Nature **317**, 505 (1985).

[34] T. W. B. Kibble, J. Phys. A: Math. Gen. **9**, 1387 (1976).

[35] C. N. Weiler, T. W. Neely., D. R. Scherer, A. S. Bradley, M. J. Davis, and B. P. Anderson, Nature **455**, 948 (2008).

[36] J. R. Anglin and W. H. Zurek, Phys. Rev. Lett. **83**, 1707 (1999).

[37] W. H. Zurek, Phys. Rev. Lett. **102**, 105702 (2009).

[38] B. Damski and W. H. Zurek, Phys. Rev. Lett. **104**, 160404 (2010).

[39] E. Witkowska, P. Deuar, M. Gajda, and K. Rzążewski, Phys. Rev. Lett. **106**, 135301 (2011).